\documentclass[acmsmall]{acmart}

\usepackage[
type={CC},
modifier={by},
version={4.0},
]{doclicense}
\newcommand{\code}[1]{\texttt{\small#1}} 
\usepackage{algorithm}
\usepackage{algorithmic}
\usepackage{longtable}
\usepackage{enumitem}
\setlist[itemize]{leftmargin=3.8mm}
\setlist[enumerate]{leftmargin=6mm}
\usepackage{booktabs}
\usepackage{url}
\usepackage{amsthm}
\theoremstyle{definition}

\usepackage{amsmath}
\usepackage{booktabs}
\usepackage{graphicx}
\usepackage{tikz}
 
\usepackage{multirow}
\usepackage{listings}
\usepackage{tcolorbox}
\usepackage{color}
\usepackage{xcolor}
\usepackage{caption}
\usepackage{subcaption}
\usepackage{longtable}
\usepackage{setspace}

 \usepackage[normalem]{ulem}

\definecolor{deepblue}{rgb}{0,0,0.5}
\definecolor{deepgreen}{rgb}{0,0.5,0}
\definecolor{deepred}{rgb}{0.6,0,0}
\definecolor{darkorange}{RGB}{255,140,0}
\definecolor{lightgray}{rgb}{0.93,0.93,0.93}
\definecolor{deepgray}{rgb}{0.25,0.25,0.25}

\newcommand*\circled[1]{\tikz[baseline=(char.base)]{
		\node[shape=circle,draw,inner sep=1pt] (char) {#1};}}

\lstdefinelanguage{JSON}{
	keywords={thoughts, command, name, args},
	ndkeywords={},
	ndkeywordstyle=\color{deepgreen}\bfseries,
	basicstyle=\scriptsize\ttfamily,
	numberstyle=\color{white},
	stepnumber=1,
	numbersep=8pt,
	showstringspaces=false,
	breaklines=true,
	frame=false,
	backgroundcolor=\color{white},
	commentstyle=\color{deepgreen},
	keywordstyle=\color{deepblue},
	stringstyle=\color{deepgreen},
	tabsize=4,
	captionpos=b,
	morecomment=[s]{/*}{*/},
	morestring=[b]",
	morestring=[d]',
	literate=
	*{[}{{{\color{deepblue}{\textbf{[}}}}}{1}
	{]}{{{\color{deepblue}{\textbf{]}}}}}{1}
	{:}{{{\color{deepblue}{\textbf{:}}}}}{1}
	{,}{{{\color{deepgreen}{,}}}}{1}
	{\{}{{{\color{deepblue}{\textbf{\{}}}}}{1}
	{\}}{{{\color{deepblue}{\textbf{\}}}}}}{1}
	{"*}{{{\color{deepgreen}{"*}}}}{1},
}

\lstdefinelanguage{TypeScript}{
    keywords={typeof, new, true, false, catch, function, return, null, catch, switch, var, if, in, while, do, else, case, break, interface, extends, implements, package, protected, private, public, export, enum, const, let, async, await, static, get, set, readonly, declare, namespace, type},
    sensitive=true,
	basicstyle=\scriptsize\ttfamily,
    keywordstyle=\bfseries\color{blue},
    comment=[l]{//},
    commentstyle=\itshape\color{green!40!black},
    morecomment=[s]{/*}{*/},
    morestring=[b]",
    morestring=[b]',
    stringstyle=\color{deepgreen},
    identifierstyle=\color{black},
    morekeywords={string, number, boolean, any, void, undefined, never, unknown, Record, Array, Promise, Symbol},
	breaklines=true,
	showstringspaces=false,
}

\lstdefinelanguage{Java}{
	basicstyle=\small\ttfamily,
	numberstyle=\color{deepgray},
	stepnumber=1,
	numbersep=8pt,
	showstringspaces=false,
	breaklines=true,
	frame=lines,
	backgroundcolor=\color{lightgray},
	commentstyle=\color{deepgreen},
	keywordstyle=\color{deepblue},
	stringstyle=\color{deepred},
	tabsize=4,
	captionpos=b,
	morekeywords={public, class, void, int, if, else, for, while, return, true, false},
	emph={String, System},
	emphstyle=\color{darkorange},
	alsoletter={.,;:[]()},
}

\lstdefinestyle{bashstyle}{
	language=bash,
	basicstyle=\ttfamily \scriptsize,       %
	backgroundcolor=\color{gray!0},  %
	breaklines=true,                  %
	keywordstyle=\color{blue}\bfseries,  %
	commentstyle=\color{green!60!black}, %
	stringstyle=\color{red!60!black}     %
}

\newcommand{\name}{Execution\-Agent}

\setcopyright{cc}
\setcctype{by}
\acmDOI{10.1145/3728922}
\acmYear{2025}
\acmJournal{PACMSE}
\acmVolume{2}
\acmNumber{ISSTA}
\acmArticle{ISSTA047}
\acmMonth{7}
\received{2025-02-27}
\received[accepted]{2025-03-31}

\begin{document}

	
\title[You Name It, I Run It: An LLM Agent to Execute Tests of Arbitrary Projects]{You Name It, I Run It:\\ An LLM Agent to Execute Tests of Arbitrary Projects}

\author{Islem Bouzenia}
\affiliation{
  \institution{University of Stuttgart}
  \country{Germany}
}
\email{fi bouzenia@esi.dz}

\author{Michael Pradel}
\affiliation{
  \institution{University of Stuttgart}
  \country{Germany}
}
\email{michael@binaervarianz.de}

\begin{abstract}

The ability to execute the test suite of a project is essential in many scenarios, e.g., to assess code quality and code coverage, to validate code changes made by developers or automated tools, and to ensure compatibility with dependencies.
Despite its importance, executing the test suite of a project can be challenging in practice because different projects use different programming languages, software ecosystems, build systems, testing frameworks, and other tools.
These challenges make it difficult to create a reliable, universal test execution method that works across different projects.
This paper presents \name{}, an automated technique that prepares scripts for building an arbitrary project from source code and running its test cases.
Inspired by the way a human developer would address this task, our approach is a large language model (LLM)-based agent that autonomously executes commands and interacts with the host system.
The agent uses meta-prompting to gather guidelines on the latest technologies related to the given project, and it iteratively refines its process based on feedback from the previous steps.
Our evaluation applies \name{} to 50 open-source projects that use 14 different programming languages and many different build and testing tools.
The approach successfully executes the test suites of 33/50 projects, while matching the test results of ground truth test suite executions with a deviation of only 7.5\%.
These results improve over the best previously available technique by 6.6x.
The costs imposed by the approach are reasonable, with an execution time of 74 minutes and LLM costs of USD~0.16, on average per project.
We envision \name{} to serve as a valuable tool for developers, automated programming tools, and researchers that need to execute tests across a wide variety of projects.
\end{abstract}


\maketitle

\section{Introduction}
\label{sec:intro}

Executing the test suite of a software project is a critical step in various activities during software development and software engineering research.
For human developers who are contributing to open-source projects, running the tests before submitting a pull request ensures that their changes do not introduce regressions.
Likewise, the increasing popularity of large language model (LLM) agents that autonomously edit a project's code~\cite{Bouzenia2024,Zhang2024a,Liu2024a,Tao2024,Yang2024a} creates a huge demand for a feedback mechanism to validate modifications~\cite{icse2025-calibration,AgenticAISE2025}, and executing the test suite provides such a mechanism.
Finally, researchers also depend on running tests, e.g., to evaluate the effectiveness of dynamic analyses~\cite{fse2022-DynaPyt} or to create benchmarks involving test execution, such as Defects4J~\cite{Just2014}, SWE-Bench~\cite{Jimenez2023}, and DyPyBench~\cite{fse2024-DyPyBench}.

Unfortunately, executing the tests of an arbitrary project is far from straightforward in practice.
Projects are developed in various programming languages and ecosystems, each with its own set of tools, dependencies, and conventions.
Complex dependencies can pose significant challenges, especially when specific versions of libraries or tools are required.
Documentation is often incomplete, inconsistent, or entirely missing, forcing developers to infer the necessary steps.
Moreover, projects may have implicit assumptions about the environment, such as operating system specifics or required system configurations, that are not explicitly stated.
The diversity of testing frameworks further complicates the process, as each framework has a unique setup and execution procedure.

Currently, there are three primary methods for executing the tests of a given project, each with notable limitations.
The first method involves manually following the project's documentation and resolving any issues through trial-and-error.
This approach is time-consuming and does not scale well with the number of projects.
The second method is to reuse existing continuous integration/continuous deployment (CI/CD) workflows employed by project maintainers.
However, not all projects have such workflows, and even when they exist, their execution often depends on a specific CI/CD platform, such as GitHub Actions, which is not fully accessible to the public.
Directly using CI/CD workflows is further complicated by the fact that there are several popular platforms, each with its own configuration scripts and technology stack.
The third method is to implement an automated, heuristic script designed to cover common cases.
Such scripts typically focus on a single programming language and ecosystem, and they lack the flexibility to handle arbitrary projects:
For example, Flapy~\cite{Gruber023} works only on Python projects that are available on PyPy, and even within the Python ecosystems fails to successfully run many test suites.
Another example is the pipeline used to create GitBug-Java~\cite{silva2024gitbug}, which, starting from 66,042 repositories, eventually managed to execute the tests of only 228 repositories.

An effective solution would need to address multiple challenges.
First, the approach should be aware of the latest technologies and tools for a range of popular programming languages.
Second, the approach should be capable of understanding incomplete and partially outdated documentation.
Third, the approach needs a way to interact with the system, such as executing commands, monitoring outputs, and handling errors.
Finally, the approach must be able to assess whether the setup process has been successful, and if not, address any problems until the test suite runs successfully.
To the best of our knowledge, there is currently no existing work that addresses these challenges.
This gap presents a substantial obstacle for developers, automated coding techniques, and researchers who need a reliable and scalable solution for running tests across a wide variety of projects.
Motivated by these challenges, the question addressed in this work is:
\emph{Given an arbitrary project, how can we automatically build the project and run its test suite?}

This paper presents \name{}, the first LLM-based agent for autonomously setting up arbitrary projects and executing their test suites.
The approach addresses the four challenges described above as follows.
For the first challenge, we present the novel concept of meta-prompting, which asks a recently trained LLM to automatically generate up-to-date, technology-specific, and programming language-specific guidelines instead of manually engineering and hard-coding a prompt.
We address the second challenge by using the LLM to parse and understand the project's documentation and web resources related to the project.
To handle the third challenge, we connect the agent to a set of tools, such as a terminal, to execute commands, monitor outputs, and interact with the system.
Finally, we address the fourth challenge by enabling the agent to iteratively refine its process based on feedback from previous steps, similar to how a human developer would work.

To evaluate \name{}, we apply it to 50 open-source projects that use 14 different programming languages and many different build and testing tools.
The approach successfully sets up and executes the test suites of 33/50 projects.
Comparing to several baselines, such as manually designed and LLM-generated scripts targeting projects in a specific language, as well as a general-purpose LLM-based agent, we find \name{} to outperform the best available technique by 6.6x.
To validate the test executions, we compare them against a manually established ground truth and find that the test results (in terms of the number of passed, failed, and skipped tests) closely resemble the ground truth, with an average deviation of 7.5\%.
Studying the costs of the approach, we find that the average execution time is 74 minutes per project, and the average LLM costs are USD~0.16 per project.
Overall, these results demonstrate that \name{} has the potential to serve as a valuable tool for developers, automated programming tools, and researchers that need to execute tests across a wide variety of projects.

In summary, this paper contributes the following:
\begin{enumerate}
	\item The first autonomous, LLM-based agent for automatically setting up arbitrary projects and executing their test suites.
	
	\item The novel concept of meta-prompting, which allows the agent to query an LLM for the latest guidelines and technologies related to building projects and running test suites.
	
	\item Several technical insights on design decisions that enable the agent to effectively interact with the system, execute commands, monitor outputs, and handle errors.
	
	\item Empirical evidence that the approach can successfully execute the test suites of a wide variety of projects, clearly outperforming existing techniques.
\end{enumerate}

\section{Approach}

The following presents our approach for executing the test of arbitrary projects.
We start by defining the problem we aim to address (Section~\ref{s:pb}), then provide an overview of our approach (Section~\ref{s:overview}), and finally describe the components of our approach in detail (Sections~\ref{s:phase1} and ~\ref{sec:loop}).

\subsection{Problem Statement}
\label{s:pb}

The problem we aim to address is the following:
Given a software project, identified, e.g., by a URL of a git repository or a file path, we want to automatically generate scripts to install the project and run its tests.
Specifically, the desired output consists of two scripts: one to create an isolated environment, such as a container, and one to build the project and run its tests within the isolated environment.
By running the tests, we mean executing the project's test suite and collecting the results, such as the number of tests that pass, fail, and are skipped.

We aim to address this problem in a way that offers two important properties.
First, the approach should be technology-agnostic, i.e., it should support projects written in different programming languages, using different build systems, and using different testing frameworks.
This property is crucial as it allows the approach to be used across a wide range of projects.
Second, the approach should be fully automated, i.e., it should not require any manual intervention or additional information beyond the project itself.
This property is essential to ensure that the approach can be used at scale and without human intervention.

The formulated problem has, to the best of our knowledge, not been addressed by prior work.
In particular, existing approaches either focus on specific programming languages or ecosystems~\cite{Gruber023,silva2024gitbug} or are based on significant manual intervention~\cite{fse2024-DyPyBench}.

\subsection{Overview of \name{}}
\label{s:overview}

\begin{figure}
	\centering
	\includegraphics[width=\linewidth]{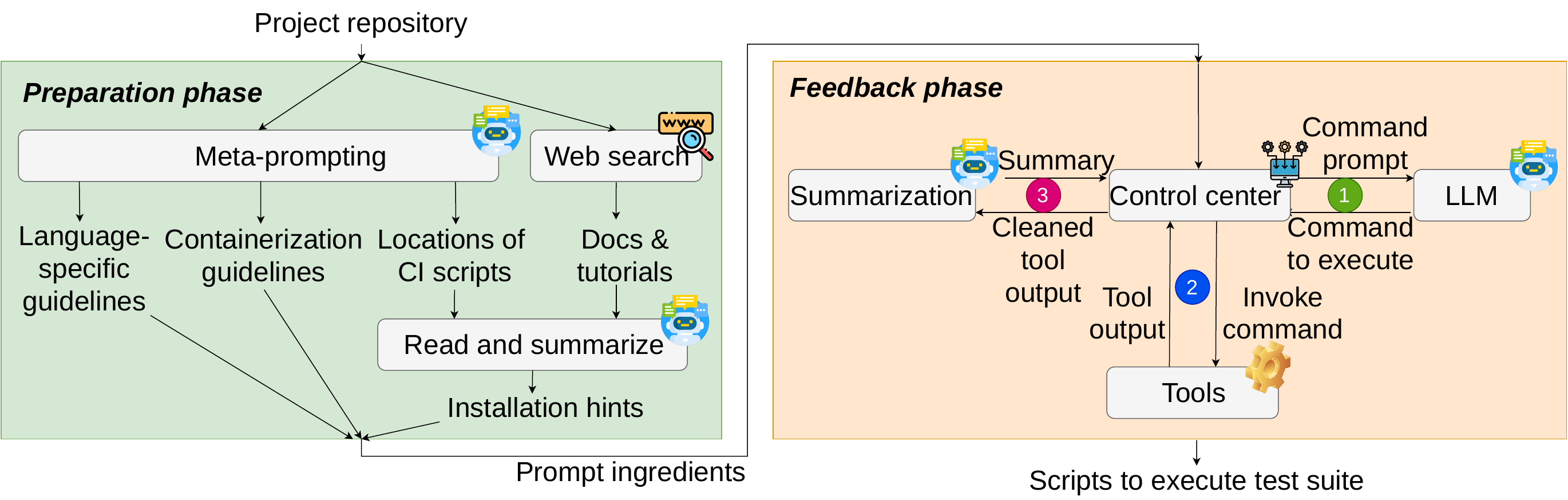}
	\caption{Overview of \name{}.} 
	\label{fig:overviewfigure}
\end{figure}

We address the above problem by presenting an approach that leverages LLMs and the concept of LLM agents.
By \emph{LLM agent} we here mean a system that uses an LLM to autonomously interact with a set of tools in order to achieve a specific goal~\cite{AgenticAISE2025}.
The tools we use are similar to those a human tasked with setting up a project might use, such as commands available via a terminal.
The intuition behind this approach is that the LLM can ``understand'' various sources of information, such as project documentation, existing scripts, and the output of tools, and use this understanding to decide on the steps necessary to execute the tests of the given project.
As shown in Figure~\ref{fig:overviewfigure}, the approach encompasses two phases, which we briefly describe in the following. 

\paragraph{Phase 1: Preparation}
Given a project repository, this phase gathers information required to construct the initial prompt of the agent, called \emph{prompt ingredients}.
A key challenge is creating these prompt ingredients in a technology-agnostic way, as the project may be written in any programming language and use any testing framework.
Our approach addresses this challenge through \emph{meta-prompting}, a novel concept that allows the agent to query an LLM for the latest guidelines and technologies related to building projects and running test suites.
Specifically, \name{} uses meta-prompting to generate language-specific guidelines, guidelines about the latest containerization technology, and possible locations of commonly used CI/CD scripts.
In addition, the approach queries the web to gather hints on building the given project. Finally, our approach uses an LLM to generate general-purpose, language-independent guidelines for creating a build and test environment for an arbitrary project. These guidelines help the agent follow a high-level plan to reach its goals.
The generated and gathered information is then passed as prompt ingredients to the second phase.

\paragraph{Phase 2: Feedback loop}
This phase repeatedly invokes tools, as guided by the LLM, in order to build the project and execute its test suite.
Specifically, \name{} repeatedly iterates through three steps.
Step~\circled{1} queries the LLM for the next command to execute, using a dynamically updated \emph{command prompt} that contains the prompt ingredients from Phase 1 and a summary of the already invoked commands.
Step~\circled{2} executes the command suggested by the LLM by invoking one of the tools.
Because the output of a tool may be verbose and contain irrelevant information, step~\circled{3} requests the LLM to summarize the output and extract the most relevant information.
The summarized output is then used to update the command prompt, and the three steps repeat until the agent determines that the test suite has been successfully executed.
This entire process is managed by a component we call the \emph{control center}.

\medskip
\noindent The steps taken by \name{} are described in more detail in Algorithm~\ref{algo:mainalgo}, which we will explain in the following sections.

\begin{algorithm}[t]
\caption{High-level algorithm of \name{}}
\label{algo:mainalgo}
\renewcommand{\algorithmicrequire}{\textbf{Input:}}
\renewcommand{\algorithmicensure}{\textbf{Output:}}
\begin{spacing}{0.95}
\fontsize{9pt}{0mm}
\begin{algorithmic}[1]
	\REQUIRE Repository URL $u$
	\ENSURE Scripts to set up an environment and run tests
	
	\STATE \texttt{// Phase 1: Preparation} \label{l:phase1 start}
	\STATE $\mathit{lang\_guidelines} \gets$ \texttt{LLM("Give language-specific guidelines", get\_languages($u$))}
	\STATE $\mathit{container\_guidelines} \gets$ \texttt{LLM("Give guidelines on recent containerization technology")}
	\STATE $\mathit{ci\_paths} \gets$ \texttt{LLM("Give common locations of CI scripts within a repo")}
	\STATE $\mathit{ci\_hints} \gets$ \texttt{read\_ci\_scripts\_and\_summarize($u$, $\mathit{ci\_paths}$)}
	\STATE $\mathit{install\_hints} \gets$ \texttt{search\_web($u$)} \label{l:phase1 end}
	
	\STATE
	\STATE \texttt{// Phase 2: Feedback loop} \label{l:phase2 start}
	\STATE $\mathit{attempts\_left} \gets 3$
	\STATE $\mathit{p\_attempt\_lessons} \gets$ ""
	\WHILE{$\mathit{attempts\_left}$}
		\STATE $\mathit{p\_cmd} \gets$ \texttt{create\_command\_prompt($\mathit{lang\_guidelines}$, $\mathit{container\_guidelines}$, $\mathit{ci\_hints}$, $\mathit{install\_hints}$, $\mathit{p\_attempts\_lessons}$),} \label{l:create prompt}
		\STATE $\mathit{budget\_left}  \gets$ 40
		\WHILE{$\mathit{budget\_left}$}
			\STATE \texttt{// Step 1: Get next command}
			\STATE $\mathit{thought}, \mathit{cmd \gets}$ \texttt{LLM($\mathit{p\_cmd}$)}
			\STATE \texttt{// Step 2: Execute command} \label{l:execute start}
			\STATE $\mathit{cmd\_result} \gets$ \texttt{execute($\mathit{cmd}$)}
			\IF{$\mathit{cmd}$ == \texttt{"task\_done"} \AND \texttt{valid($\mathit{cmd\_result}$)}}
			\STATE $\mathit{scripts} \gets$ \texttt{read\_target\_files()}
			\RETURN $\mathit{scripts}$ \label{l:task done}
			\ENDIF \label{l:execute end}
			\STATE \texttt{// Step 3: Summarize command output}
			\STATE $\mathit{p\_summarize} \gets$ \texttt{create\_prompt($\mathit{p\_cmd}$, $\mathit{cmd\_result}$)} \label{l:summarize start}
			\STATE $\mathit{cmd\_result\_summary} \gets$ \texttt{LLM($\mathit{p\_summarize}$)} \label{l:summarize end}
			\STATE $\mathit{p\_cmd} \gets$ \texttt{update\_prompt($\mathit{p\_cmd}$, $\mathit{thought}$, $\mathit{cmd}$, $\mathit{cmd\_result\_summary}$)} \label{l:update prompt}
			\STATE $\mathit{budget\_left} \gets \mathit{budget\_left} - 1$
		\ENDWHILE
		\STATE $\mathit{attempts\_left} \gets \mathit{attempts\_left} - 1$
		\STATE $\mathit{p\_attempt\_lessons} \gets analyze\_commands\_and\_thoughts(cmd\_list, thoughts\_list)$ \label{l:lessons}
	\ENDWHILE \label{l:phase2 end}
\end{algorithmic}
\end{spacing}
\end{algorithm}

\subsection{Preparation Phase}
\label{s:phase1}

The preparation phase is motivated by two goals:
(1) to gather project-specific information that could be helpful for building the given project and running its tests, and
(2) to obtain guidelines that will help the LLM agent invoke the right tools and commands during the feedback loop in phase~2.
To reach these goals, we follow a set of steps that are detailed in Algorithm~\ref{algo:mainalgo} between lines~\ref{l:phase1 start} and~\ref{l:phase1 end}, and which we present in detail in the following. 
Since several of these steps use the idea of meta-prompting, we begin by explaining this concept.

\subsubsection{Meta-Prompting}
A key challenge faced by \name{} is that technologies and best practices for building software projects and running test suites are constantly evolving.
One way to address this challenge would be to spend significant amounts of time in engineering suitable prompts for the LLM agent.
Such prompts should provide guidelines on how to use common build tools, package managers, testing frameworks, etc.\ for all the programming languages, software ecosystems, and tools that the agent might encounter.
However, this approach comes with several drawbacks: it is time-consuming, unlikely to cover all relevant technologies, and may become outdated quickly.

Instead of engineering such prompts manually, we leverage the impressive knowledge of LLMs to generate prompts with up-to-date guidelines targeted at the given project.
Since LLMs are often trained on a vast range of documents, including documents describing the latest technological developments, they offer a built-in mechanism for adapting to evolving technology landscapes.
We leverage this capability by using a high-level prompt, called a \emph{meta-prompt}, to query the LLM for technology-specific guidelines and other information, which we then use to construct the command prompt for the agent.
Specifically, \name{} uses meta-prompting to generate three kinds of prompt ingredients, which we describe in the following.

\subsubsection{Language-Specific and Language-Independent Guidelines}  
During initial experiments, we observed that the LLM agent benefits from guidelines that describe how to typically build projects written in a specific language and how to run their tests.  
Adopting the idea of meta-prompting, \name{} creates such guidelines by querying an up-to-date LLM to generate a list of language-specific guidelines. To this end, we heuristically infer the main programming language of the project during the preparation phase (using GitHub's Linguist library\footnote{\url{https://github.com/github-linguist/linguist}}) and use this information to generate the language-specific guidelines.

\begin{figure}[t]
	\centering
	\begin{tcolorbox}[colframe=blue!50!green, colback=blue!5!white, coltitle=white, title=Guidelines Meta Prompt (for Java), 
		fonttitle=\bfseries \footnotesize, rounded corners=southwest, width=1\textwidth, boxrule=0.25pt, 
		enlarge left by=0.25em, left=0.25pt, right=0.25pt, top=0.25pt, bottom=0.25pt]
		\sffamily \footnotesize I want to install and run tests of  a \textbf{Java} project. Give a list of 10 guidelines that I should follow to achieve my task (e.g., to figure out installation steps, dependencies, how to run tests) while considering latest practices and technologies.
	\end{tcolorbox}
	
	\vspace{0.5em} %
	
	\begin{tcolorbox}[colframe=green!50!black, colback=green!5!white, coltitle=white, title=LLM Response, 
		fonttitle=\bfseries \footnotesize, rounded corners=southeast, width=1\textwidth, boxrule=0.25pt, 
		enlarge left by=0.25em, left=0.25pt, right=0.25pt, top=0.25pt, bottom=0.25pt]
		\sffamily \footnotesize 1. Review Project Documentation and README
		
		Why: Documentation typically includes installation steps, dependencies, and setup instructions.
		
		What to do: Check the README.md or other documentation files for prerequisites and specific instructions for installation and testing.
		
		\vspace{0.5em}
		
		2. Verify Java Version Compatibility
		
		Why: Java projects may target specific versions (...)
		
		What to do: Confirm the Java version required by checking the pom.xml (Maven) or build.gradle (Gradle) file (...)
		
		\vspace{0.5em}
		3. Check for Dependency Management Tool
		
		(...)
		\vspace{0.5em}
		
		10. Generate Test Reports for Results and Debugging	
		
		Why: Test reports provide detailed logs and summaries, aiding 
		in debugging.
		
		What to do: Maven projects can use mvn surefire-report:report to generate reports. Gradle usually generates HTML reports by default (build/reports/tests/test/index.html).
		
	\end{tcolorbox}
	
	\caption{Meta-prompting to obtain guidelines for Java projects. Response shortened for brevity.}
	\label{fig:prompt_response}
\end{figure}

Figure~\ref{fig:prompt_response} illustrates an example of a meta-prompt for a Java project and the response from the LLM.
The prompt asks for a list of ten guidelines for a specific programming language.
The reason for asking for a specific number of guidelines is that otherwise, the LLM sometimes generates only a few broad guidelines, which are lacking in detail, or may generate an extensively long list, which is hard to process.
To avoid such variation, we explicitly request a specific number, and we found ten to be sufficiently comprehensive.  

In addition to creating language-specific guidelines, we instruct the agent to collect further information about the projects and tests using a set of language-independent guidelines that we also generate with an LLM. These language-independent guidelines ensure that our approach remains effective even when a project's setup does not strictly adhere to the common practices of its main language, for instance, due to significant influence from secondary languages.

\subsubsection{Container Guidelines}
To ensure that the project's tests are executed in an isolated environment, \name{} aims at building and testing the project inside a container.
Similar to the language-specific guidelines, we use meta-prompting to query the LLM for guidelines on the latest and most used containerization technology.
For example, at the moment of writing, the GPT-4o model responds to our meta-prompt by stating that Docker is the best option.

\subsubsection{Existing CI/CD Scripts}

\begin{figure}[t]
	\centering
	\begin{tcolorbox}[colframe=white!70!black, colback=gray!0, coltitle=black, title=Summarization response, 
		fonttitle=\bfseries \footnotesize, rounded corners=southwest, width=1\textwidth, boxrule=0.75pt, 
		enlarge left by=-0.0em, left=0pt, right=0pt, top=0pt, bottom=2pt]
\begin{lstlisting}[language=JSON]
{ 'summary': 'The karma.conf.js file configures the testing framework for Bootstrap, utilizing Jasmine for unit testing and Rollup for preprocessing. It supports headless testing with Chrome and Firefox, and includes configurations for coverage reporting and BrowserStack integration.',
  'extracted dependencies': ['jasmine', 'karma', 'karma-jasmine','karma-rollup-preprocessor', 'karma-chrome-launcher', 'karma-firefox-launcher', 'karma-detect-browsers', 'karma-coverage-istanbul-reporter'],
  'important commands': ['npm test'],
  'important files/links/hyperlinks': [] }
\end{lstlisting}
	\end{tcolorbox}
\caption{Example of an LLM-generated summary.}
\label{fig:summary_instruction}
\end{figure}

Some projects include CI/CD scripts that provide hints about installation steps and project dependencies. To leverage such scripts, \name{} follows a three-step process. First, it uses meta-prompting to query the LLM for common folder names, file names, and extensions associated with CI/CD scripts.  
Second, based on the obtained list, it searches the repository for matching files and folders.  
Third, the identified files are analyzed by the LLM to determine which files are relevant to building and testing.  

While CI/CD scripts can sometimes be outdated or incomplete, they may still provide useful insights into the overall setup process and any required non-standard steps.  
To extract the most relevant information, \name{} queries an LLM with the raw file contents, asking it to provide:  
(1) a concise natural language summary of the script,  
(2) a list of project dependencies,  
(3) important commands for setting up and testing, and  
(4) references to relevant files or links. Figure~\ref{fig:summary_instruction} shows an example of the LLM's response to the summarization prompt.

\subsubsection{Web Search}

Some projects maintain explicit and structured documentation on external platforms, such as \textit{readthedocs.com}, which can provide clearer installation instructions than their respective GitHub repositories. Additionally, search engines efficiently retrieve the most relevant resources based on a few keywords, reducing the need for exhaustive exploration.

To leverage web resources, \name{} queries a search engine with ``How to install the <LANGUAGE> project '<PROJECT>' from source code?'', where <LANGUAGE> and <PROJECT> correspond to the project's primary programming language and name, respectively. Instead of parsing webpages sequentially—which would increase the number of queries and associated costs—the approach extracts text from the top five results and asks an LLM to summarize them. The summarization process extracts relevant build and test dependencies, as well as installation steps, including commands and configuration details. If a search result lacks useful installation hints, the model is instructed to return: “This page does not contain setup-related information.”

\subsection{Feedback Phase}
\label{sec:loop}
Based on the prompt ingredients obtained in the preparation phase, the second phase of our approach is a feedback phase that iteratively invokes tools to build the project and execute its test suite.
As shown in lines~\ref{l:phase2 start} to~\ref{l:phase2 end} of Algorithm~\ref{algo:mainalgo}, the feedback phase consists of two nested loops.
The inner loop runs a series of commands, guided by the LLM, until either the test suite completes successfully or a configurable command limit (default: 40) is reached. The outer loop allows for multiple attempts to build and execute the test suite, with a configurable maximum number of attempts (default: 3). This design allows recovery from occasional failures where the build process encounters an error, preventing the tests from running. By allowing multiple attempts, the system can adjust its approach to overcome such errors.
At the start of each outer loop iteration, the system incorporates lessons learned from the previous attempt (if any) into the command prompt. These lessons are generated by prompting the LLM to analyze the previous sequence of thoughts and commands, identify challenges encountered, and suggest adjustments for the next iteration (line~\ref{l:lessons}).
Each iteration of the inner loop consists of three steps, as outlined in Figure~\ref{fig:overviewfigure} and detailed in the following.

\subsubsection{Step 1: LLM Selects the Next Command}

The core of our approach is an LLM that selects the next command to execute based on the current state of the installation process.
To guide the LLM in selecting the next command, the approach constructs a \emph{command prompt} that contains the prompt ingredients obtained in the preparation phase, as well as a summary of the commands executed so far.
The command prompt consists of predefined static sections, and dynamic sections, which are either created based on the preparation phase (line~\ref{l:create prompt}) or updated at the end of each iteration of the inner loop (line~\ref{l:update prompt}).
Specifically, the prompt contains the following sections:
\begin{enumerate} 
	
	\item Agent role (static): Defines the primary task of the agent and the success criteria. Concretely, the agent's role is to build the given project and ensure that the test cases execute properly.
	
	\item Goals (static): Outlines the specific objectives the agent must accomplish, namely:
	\begin{itemize}
		\item Gather installation-related information and requirements.
		\item Write scripts to build the project and execute its tests.
		\item Run the produced scripts and refine them, if necessary.
		\item Analyze the test execution results and summarize them by providing the number of passing, failing, and skipped tests.
	\end{itemize}
	
	\item Tools (static): Lists the tools available to the agent. We describe the tools in detail in Section~\ref{s:tools}.
		
	\item Guidelines (dynamic): Contains the programming language-specific guidelines and the containerization guidelines obtained in the preparation phase.
	
	\item Installation hints (dynamic): Contains the installation hints obtained from any existing CI/CD scripts and the web search. It also contains lessons learned from previous iterations of the outer feedback loop.
	
	\item Tool invocation history (dynamic): Summarizes the tools invoked so far, as well as the output that was produced.
	Initially, this section is empty.
	
	\item Instruction for creating the next command prompt (static):
	Specifies that the LLM should provide the next command to execute based on the context available in the prompt.
	We define the expected format of the LLM's response in a TypeScript interface, as shown in Figure~\ref{fig:command_instruction}.
	The expected format includes the agent's thoughts, as well as the name of the tool to invoke and the arguments to pass to the tool.
	The rationale for asking the LLM to describe its thoughts is two-fold:
	First, it has been shown empirically to improve the quality of the LLM's responses~\cite{wei2022chain}.
	Second, it allows for easier debugging and understanding of the LLM's decisions.
	If the LLM fails to respond in this format, the control center returns an error message to the LLM, mentioning the usage of the wrong format and reminding the LLM of the correct one.
\end{enumerate}

\begin{figure}[t]
	\centering
	\begin{tcolorbox}[colframe=white!70!black, colback=gray!0, coltitle=black, title=Test results, 
		fonttitle=\bfseries \footnotesize, rounded corners=southwest, width=1\textwidth, boxrule=0.25pt, 
		enlarge left by=-0.0em, left=0pt, right=0pt, top=0pt, bottom=0.0pt]
\begin{lstlisting}[language=TypeScript]
interface Response {// Express your thoughts based on the information that you have collected so far, the possible steps that you could do next, and also your reasoning.
thoughts: string;
tool: {name: string; args: Record<string, any>; }; }

Here is an example of a command call that you can output:
{ "thoughts": "I need to check the files and folders available within this repository.",
"tool": {"name": "linux_terminal", "args": {"cmd": "ls"} } }
\end{lstlisting}
	\end{tcolorbox}
	\caption{TypeScript interface to specify the JSON format of the LLM's response to the command prompt.}
	\label{fig:command_instruction}
\end{figure}

\subsubsection{Step 2: Invoking Tools}
\label{s:tools}

To enable our approach to take the steps necessary for building the project and running its tests, we provide four tools to the agent.
\name{} invokes these tools based on the LLM's response to the command prompt (lines~\ref{l:execute start} to~\ref{l:execute end}).

\paragraph{Terminal}

Similar to human developers, access to a terminal is crucial to successfully set up a project and run its tests, e.g., to build dependencies or list available files.
We provide the agent with the capability to execute any command available in a Linux terminal via the \code{linux\_terminal} tool.
The tool takes a command to execute as input and returns the output of the command.

\paragraph{File I/O Tools}

While the terminal would, in principle, be sufficient to perform any kind of operation on the system, we provide two additional tools for interacting with the file system.
These tools are meant to emulate the developer's interaction with text editors, where they open a file, read parts of it, and sometimes make changes by writing to it.
The \code{read\_file} tool takes a file path as input and returns the content of the file.
The \code{write\_file} tool takes a file path and content as input and writes the content to the file.
The latter tool is particularly useful for writing the scripts expected as the output of \name{}, e.g., a \code{Dockerfile} that creates a container and a \code{install.sh} script with the sequence of commands to build dependencies, compile or build the project, and eventually run the test suite.

\paragraph{End of Task}

In addition to the tools described above, we provide the agent with a special tool called \code{task\_done}.
This tool is used to signal that the agent has successfully completed its task, i.e., that the project has been built and its tests have been executed.
The tool expects a natural language description that explains why the agent has finished the task.

\subsubsection{Step 3: Summarization and Extraction}

The output of tools can be verbose and contain lots of irrelevant information.
For example, suppose the agent executes a command to list all files in a directory or reads the content of a file, then simply returning the output of the tool to the LLM for the next command would quickly exceed the available prompt size.
Even with LLMs that offer a large prompt size, it is beneficial to reduce the amount of information in the prompt to keep the agent focused on the most important information and to reduce the overall costs of the approach.

To reduce the amount of text that results from a tool invocation, \name{} shortens the output whenever the output exceeds a certain number of tokens (default: 200).
The approach asks another LLM to summarize the output and extract the most relevant information (lines~\ref{l:summarize start} to~\ref{l:summarize end}).
The expected format for the summary is the same as illustrated in Figure~\ref{fig:summary_instruction}.
Once the output of the most recently invoked command has been summarized, the approach appends the command and its output summary to the tool invocation history section of the command prompt (line~\ref{l:update prompt}).

\subsubsection{Control Center}

The control center orchestrates the three steps described above, managing the interaction between the LLMs and the tools. Specifically, it performs the following tasks:  

\begin{itemize}  
	\item Parse the LLM output and validate whether it conforms to the specified output format.   
	
	\item Invoke the next step as specified in Algorithm~\ref{algo:mainalgo}, e.g., by executing a specific command.  
	
	\item Monitor tool execution by tracking launched processes within the container and detecting when they terminate.  
	Since some commands may take a long time before producing any output, merely watching stdout and stderr is insufficient.  
	Instead, the control center ensures that process completion is explicitly detected.  
	The complete stdout and stderr outputs are returned to the LLM once the process terminates.  
	
	\item Enforce a configurable timeout (default: five minutes) for any command execution.  
	If a command does not finish within this interval, the control center provides the LLM agent with the available output and process status.  
	The LLM then decides among three options:  
	(1) wait for another five minutes,  
	(2) provide input to the running tool (useful for interactive prompts, such as confirming an installation with ``y''),  
	or (3) terminate the command.  
	
	\item Clean the output of tools by removing terminal color codes and other special characters, such as those used for progress bars.  
\end{itemize}

When the agent selects the \code{task\_done} command, it indicates that the agent believes that the project has been successfully built and that its test suite has been executed.
Before terminating \name{}, the control center validates that the agent has indeed met all goals.
This involves checking whether three files exist: a file to create a container (e.g., a \code{Dockerfile}), an installation script to be executed within the container, and a file that contains the test results (i.e., the number of passed, failed, and skipped tests).
If all goals are met, \name{} returns the installation and test scripts to the user (line~\ref{l:task done}).
Otherwise, the control center provides feedback to the LLM agent, pointing out what exactly is missing, and asks for a new command. Figure~\ref{fig:bash_scripts} gives an example of the scripts produced for the OpenVPN project. The two scripts show the configuration chosen by \name{}, the installed dependencies, and the steps to build the project and execute its tests.

\begin{figure}[t]
	\centering
	
	\begin{minipage}{1\textwidth}
		\begin{tcolorbox}[colframe=white!70!black, colback=gray!0, coltitle=black, title=Dockerfile, 
			fonttitle=\bfseries \footnotesize, rounded corners=southwest, width=\textwidth, boxrule=0.75pt, 
			enlarge left by=-0.0em, left=0pt, right=0pt, top=0pt, bottom=0pt]
			\lstset{style=bashstyle}
\begin{spacing}{.9}
\begin{lstlisting}
FROM ubuntu:20.04
# Set the timezone to avoid interactive prompts
ENV TZ=XXX/YYY # omitted for anonymity
RUN ln -snf /usr/share/zoneinfo/$TZ /etc/localtime && echo $TZ > /etc/timezone
# Update package lists and install necessary build tools and dependencies
RUN apt-get update && \
apt-get install -y --no-install-recommends \
sudo \
build-essential \
liblzo2-dev \
libpam0g-dev \
liblz4-dev \
libcap-ng-dev \
libnl-genl-3-dev \
linux-libc-dev \
autoconf \
automake \
libtool \
cmocka-dev \
python3-docutils \
git \
&& apt-get clean && rm -rf /var/lib/apt/lists/* || exit 0
WORKDIR /app
# Clone the OpenVPN repository
RUN git clone https://github.com/OpenVPN/openvpn.git || exit 0
WORKDIR /app/openvpn
CMD ["bash"]
\end{lstlisting}
\end{spacing}
\vspace{-.5em}
\end{tcolorbox}
	\end{minipage}
	\hfill
	\begin{minipage}{1\textwidth}
		\begin{tcolorbox}[colframe=white!70!black, colback=gray!0, coltitle=black, title=install\_and\_run.sh, 
			fonttitle=\bfseries \footnotesize, rounded corners=southeast, width=\textwidth, boxrule=0.75pt, 
			enlarge left by=-0.0em, left=0pt, right=0pt, top=0pt, bottom=2pt]
			\lstset{style=bashstyle}
\begin{spacing}{.9}
\begin{lstlisting}
#!/bin/bash
# Run autoconf to prepare the build system
autoreconf -fvi
# Configure the build system
./configure --enable-werror
# Compile the project
make -j$(nproc)
# Run tests
make check VERBOSE=1
echo "Setup and testing complete."
			\end{lstlisting}
\end{spacing}
\vspace{-.5em}
		\end{tcolorbox}
	\end{minipage}
	
	\caption{Scripts produced by \name{} for the OpenVPN project.}
	\label{fig:bash_scripts}
\end{figure}

\section{Evaluation}

Our evaluation aims to answer to following research questions:
\begin{itemize}
	\item RQ1 (effectiveness): How effective is \name{} at correctly setting up projects and running their test cases?
	
	\item RQ2 (costs): What are the costs of \name{} in terms of execution time and token usage when interacting with the LLM?
	
	\item RQ3 (ablation study): What is the impact of different components and configurations?
	
	\item RQ4 (trajectories): How does \name{} interact with the tools, and what trajectories does it take to reach the goal? 
\end{itemize}

\subsection{Experimental Setup}

\subsubsection{Implementation and Model}

\name{} is implemented in Python and bash.
To allow for an isolated execution of \name{} itself, we use a Docker container.
The agent is based on OpenAI's GPT-4o-mini model, which we access through their Python API. The usage of other models is possible as long as they support JSON output format specified through an interface. At the moment of writing the paper, Claude, DeepSeek, and Llama models support this feature.

\subsubsection{Metrics}
\label{s:metrics}

The task performed by \name{} consists of two sub-tasks: (i) building the project, and (ii) running the test suite.
We measure how effective the approach is at performing these tasks by measuring the \emph{successful build rate}, i.e., the proportion of projects where the approach manages to correctly build the project, and the \emph{successful testing rate}, i.e., the proportion of projects where the approach manages to run the test suite.
To determine these rates, we manually inspect the produced scripts and the test results.

To better understand the output of \name{}, we measure the \emph{script size} of the produced scripts, which we define as the number of commands, excluding any comments, white spaces, and ``echo'' commands.
We split composite commands and count their components individually.
For example, the line  \code{make . \&\& make test} is counted as two commands \code{make .} and \code{make test}.

To validate the test suite executions, we compare the test execution results produced by \name{} with a manually established ground truth.
The ground truth is obtained by searching for CI/CD logs of the targeted projects, and by extracting the number of passed, failed, and skipped tests.
We compare the results of \name{} with the ground truth in terms of these three numbers.
Specifically, we compute the \emph{deviation from ground truth} as $\mathit{deviation} = |\frac{(nb_{EA} - nb_{GT})}{nb_{GT}}| * 100$, where $nb_{EA}$ is the number of tests produced by \name{} and $nb_{GT}$ is the number of tests in the ground truth.
For example, if among the tests executed by \name{}, there are 95 passing tests, while the ground truth has 100 passing tests, then the deviation is 5\%.
We compute a separate deviation for the number of passing, failing, and skipped tests, and then average these deviations to obtain the overall deviation.

\subsubsection{Dataset}

To construct a dataset for evaluating \name{}, we gather 50 open-source projects from GitHub.
The following describes the selection criteria and process.

\paragraph{Criteria for selecting projects}
We select projects that meet the following criteria:
(i) \emph{Diversity in programming languages:} We select ten projects each for Python, Java, C, C++, and JavaScript, ensuring that each project has one of these languages as either the dominant or second most-used language.
(ii) \emph{Availability of ground truth test execution results:} We ensure that test execution results are accessible through logs from CI/CD platforms.
As different projects use different platforms, we collect ground truth data from GitHub Actions, CircleCI, Jenkins, CirrusCI, and occasionally, from a project's website (e.g., for TensorFlow). Our selection process prioritizes projects that maintain publicly accessible, detailed reports of test execution results, such as build/test logs on CI/CD platforms. One indicator of this is the presence of a CI/CD badge in the repository's README.
(iii) \emph{Project maturity and activity:} Each selected project has at least 100 stars and 100 commits, with at least one commit made within the six months prior to data collection. This criterion ensures that the repositories are active and maintained.

\paragraph{Dataset construction process}
Our dataset construction follows a systematic process to ensure quality and relevance. We search for the most popular projects for each of the five primary languages and rank them based on their popularity (number of stars and forks).
We then manually inspect projects, starting at the top of the ranked list, to check whether they meet the selection criteria described above.
If a project does not meet the criteria, we move to the next project in the list.
We continue this process until we have collected 50 projects that meet the criteria.

\paragraph{Dataset statistics}
The projects in our dataset exhibit significant activity and complexity. The median project has approximately 10K commits, 45K stars, and 1.4K test cases.
The code in the 50 selected projects is written in 14 programming languages.
When considering only test code, the projects cover 8 languages, including Kotlin, Scala, and TypeScript, in addition to the five primary languages. On average, 87\% of the tests are written in the main language of the project. Some projects write test cases in different languages than their main language, though.
For example, the CPython project has many test cases written in Python that test code implemented in C.

From another perspective, 38 out of the 50 projects use automated build tools or provide configuration files for them.

All projects in the dataset are capable of running on a Linux system, as confirmed by logs retrieved from CI/CD results. The dataset also highlights variations in CI/CD usage: 33 out of 50 projects contain scripts for both building and testing, 10 projects have build scripts only, and 7 projects lack both build and test scripts. Additionally, 9 projects contain submodules, indicating a higher degree of structural complexity. Table~\ref{tab:projects_details} lists all 50 projects and  their characteristics.

\begin{table*}
	\centering
	\vspace{0.3em}
	\caption{Projects used for the evaluation. Languages: J=Java, P=Python, JS= JavaScript, TS= TypeScript, KT=Kotlin, ASM=Assembly, SH=Shell. Symbols: + means successfully built but tests not executed, - means failed to build.}
	\vspace{-0.8em}
	\renewcommand{\arraystretch}{.88} %
	\resizebox{0.92\textwidth}{!}{\begin{tabular}{@{}llrrrr|rrrr@{}}
			\toprule
			\textbf{Project} & \textbf{Languages} & \multicolumn{4}{c}{\textbf{Ground Truth}} & \multicolumn{4}{c}{\textbf{ExecutionAgent}} \\
			\cmidrule{3-6} \cmidrule{7-10}
			& & \textbf{Tests} & \textbf{Pass} & \textbf{Fail} & \textbf{Skip} & \textbf{Tests} & \textbf{Pass} & \textbf{Fail} & \textbf{Skip} \\
			\midrule
			Activiti & J & 3269 & 3266 & 0 & 3 & 2392 & 2389 & 1 & 2 \\
			ansible & P & 1717 & 1703 & 0 & 14 & + & + & + & + \\
			axios & JS, TS & 203 & 203 & 0 & 0 & 203 & 195 & 8 & 0 \\
			bootstrap & JS, Html/css & 808 & 808 & 0 & 0 & 808 & 793 & 15 & 0 \\
			ccache & C, SH & 47 & 36 & 0 & 11 & 47 & 36 & 0 & 11 \\
			Chart.js & JS, TS & 3298 & 3298 & 0 & 0 & + & + & + & + \\
			commons-csv & J & 856 & 845 & 0 & 11 & 856 & 845 & 0 & 11 \\
			cpython & P, C, C++ & 478 & 462 & 0 & 16 & 478 & 433 & 4 & 42 \\
			deno & JS, TS, Rust & 1458 & 1446 & 0 & 12 & 511 & 474 & 36 & 1 \\
			distcc & C, P & 57 & 57 & 0 & 0 & 57 & 53 & 4 & 0 \\
			django & P & 17604 & 16151 & 5 & 1448 & 17605 & 16148 & 6 & 1451 \\
			dubbo & J & 14020 & 13678 & 0 & 342 & 14020 & 13678 & 0 & 342 \\
			express & JS & 1228 & 1228 & 0 & 0 & 1228 & 1228 & 0 & 0 \\
			flask & P & 484 & 481 & 0 & 3 & 484 & 480 & 1 & 3 \\
			flink & J, Scala, P & 105242 & 100644 & 0 & 4598 & + & + & + & + \\
			folly & C++, P & 3089 & 3088 & 0 & 0 & - & - & - & - \\
			FreeRTOS & C, ASM & 215 & 215 & 0 & 0 & - & - & - & - \\
			git & C, SH, Perl & 31715 & 31715 & 0 & 0 & 30264 & 29203 & 265 & 796 \\
			guava & J & 857221 & 857221 & 0 & 0 & 857221 & 857221 & 0 & 0 \\
			imgui & C++, C & 348 & 348 & 0 & 0 & - & - & - & - \\
			json & C++ & 93 & 93 & 0 & 0 & 93 & 93 & 0 & 0 \\
			json-c & C & 25 & 25 & 0 & 0 & 25 & 25 & 0 & 0 \\
			keras & P & 12110 & 11860 & 0 & 250 & 12775 & 12445 & 10 & 320  \\
			langchain & P & 1358 & 887 & 0 & 471 & + & + & + & + \\
			libevent & C & 68 & 68 & 0 & 0 & 68 & 68 & 0 & 0 \\
			mermaid & JS, TS & 3106 & 3104 & 0 & 2 & 3287 & 3276 & 0 & 11 \\
			mpv-player & C & 14 & 14 & 0 & 0 & - & - & - & - \\
			msgpack-c & C & 41 & 41 & 0 & 0 & - & - & - & - \\
			mybatis & J & 1870 & 1851 & 0 & 19 & 1870 & 1851 & 0 & 19 \\
			nest & JS & 1655 & 1655 & 0 & 0 & 1655 & 1655 & 0 & 0 \\
			node & JS, C++, P & 4218 & 4218 & 0 & 0 & - & - & - & - \\
			numpy & P, C, C++ & 44861 & 43197 & 0 & 1629 & - & - & - & - \\
			opencv & C++, C, P & 232 & 216 & 0 & 16 & 232 & 216 & 0 & 16 \\
			openvpn & C & 96 & 95 & 0 & 1 & 91 & 85 & 2 & 4 \\
			pandas & P, C & 199448 & 172973 & 1025 & 25486 & 183384 & 171565 & 2233 & 9586 \\
			pytest & P & 3762 & 3640 & 11 & 111 & 3805 & 3775 & 11 & 119 \\
			react & JS, TS, Rust & 5509 & 5509 & 0 & 0 & 9764 & 9738 & 1 & 25 \\
			react-native & C++, J, JS & 158 & 158 & 0 & 0 & 158 & 158 & 0 & 0 \\
			rocketmq & J & 2336 & 2324 & 0 & 12 & 2336 & 2324 & 0 & 12 \\
			RxJava & J & 11863 & 9404 & 0 & 2459 & 11863 & 9395 & 9 & 2459 \\
			scikit-learn & P, C, C++ & 37064 & 31900 & 125 & 5039 & 37729 & 32546 & 128 & 5055 \\
			scipy & P, C, C++ & 52672 & 49745 & 0 & 2767 & + & + & + & + \\
			spring & J, KT & 117 & 74 & 0 & 43 & 117 & 74 & 0 & 43 \\
			tensorflow & C++, P & 3148 & 0 & 0 & 0 & + & + & + & + \\
			TypeScript & JS & 96598 & 96598 & 0 & 0 & 96598 & 96598 & 0 & 0 \\
			vue & TS, JS & 1449 & 1449 & 0 & 0 & 1449 & 1449 & 0 & 0 \\
			webpack & JS & 28935 & 28815 & 0 & 120 & - & - & - & - \\
			webview & C++, C, P & 12 & 12 & 0 & 0 & - & - & - & - \\
			xgboost & C++, P, R & 208 & 207 & 0 & 1 & + & + & + & + \\
			xrdp & C, C++ & 206 & 206 & 0 & 0 & + & + & + & + \\
			\bottomrule
	\end{tabular}}
	\label{tab:projects_details}
\end{table*}

\subsubsection{Baselines}
To the best of our knowledge, there is no existing technique that addresses the same problem as \name{}.
However, we compare our approach to three related baselines, which represent the state of the art in this domain.

\paragraph{LLM scripts} LLMs have seen large amounts of data, including documentation on how to install projects. We leverage this feature to ask an LLM to generate a script that prepares an environment to build an arbitrary project and run its tests.
As our dataset is gathered by focusing on five main languages, we ask the LLM to create one specialized script for each of these five languages.
The prompt we use is given in Figure~\ref{fig:llm-prompting}.
Given the script created in response to this prompt, we attempt to build a project using the script corresponding to the project's main language.
	
\paragraph{AutoGPT} Instead of creating a specialized agent for the task of automatically setting up arbitrary projects, one could also use a general-purpose LLM-based agent.
	We compare against such an agent, AutoGPT\footnote{\url{https://github.com/Significant-Gravitas/AutoGPT}}~\cite{yang2023auto}, which, given a task description, autonomously reasons about a task, makes a plan, executes, and updates the plan over multiple iterations. Similar to \name{}, AutoGPT may call tools, such as web search, reading and writing files, and executing Python code.
	As a task description, we provide the input shown in Figure~\ref{fig:autogpt_input}.
	We use the same model and provide the same budget (maximum number of iterations) to AutoGPT as for \name{}.
	
\paragraph{Flapy} This baseline is a human-written script written to automatically set up arbitrary Python projects and run their test cases.\footnote{\url{https://github.com/se2p/FlaPy}}
	The script was originally developed as part of a study of test flakiness~\cite{Gruber023}, and by design, is limited to Python projects.

\begin{figure}
	\centering
	\begin{tcolorbox}[colframe=blue!50!green, colback=blue!5!white, coltitle=white, title=Prompt for LLM scripts baseline, 
		fonttitle=\bfseries \footnotesize, rounded corners=southwest, width=1\textwidth, boxrule=0.25pt, 
		enlarge left by=0.25em, left=0.25pt, right=0.25pt, top=0.25pt, bottom=0.25pt]
		\sffamily \footnotesize 
		
	Create a script that automatically installs a <LANGUAGE> project (on an Ubuntu Linux machine) from source code and runs test cases. The script should account for differences between projects, test frameworks, and dependencies installation. The script should be as general as possible, but should also handle special cases that you are aware of.
		
	\end{tcolorbox}
\vspace{-0.7em}
	\caption{Prompt to generate general-purpose installation scripts (``LLM scripts'' baseline).}
	\label{fig:llm-prompting}
\end{figure}

\begin{figure}
	\centering
	\begin{tcolorbox}[colframe=blue!50!green, colback=blue!5!white, coltitle=white, title=Input to AutoGPT baseline, 
		fonttitle=\bfseries \footnotesize, rounded corners=southwest, width=1\textwidth, boxrule=0.25pt, 
		enlarge left by=0.25em, left=0.25pt, right=0.25pt, top=0.25pt, bottom=0.25pt]
		\sffamily \footnotesize 
		
		You are an AI assistant specialized in automatically setting up a given project and running its test cases. For your task, you must fulfill the following goals:
		
		1. Gather installation-related information and requirements for the project <GITHUB URL>
		
		2. Write a bash script (.sh) that allows to install dependencies, prepare the environment, and launch test case execution.
		
		3. Refine the script if necessary: If an error happens or the output is not expected, refine the script.
		
		4. Once the script launches the test suite successfully, analyze the results of running the test suite to further check whether there are any major problems (for example, some test cases would fail because the project or environment is not well configured, which would mean that the previous goals were not achieved).
		
	\end{tcolorbox}	
	\vspace{-0.7em}
	\caption{Input given to the AutoGPT baseline.}
	\label{fig:autogpt_input}
\end{figure}

\subsection{Effectiveness}
\paragraph{\name{}}
The results of applying \name{} to the 50 projects are reported in Table~\ref{tab:projects_details} and summarized in Table~\ref{tab:comparison}.
\name{} successfully builds and tests 33 out of 50 projects.
Out of the 33 successfully tested projects, \name{} achieves results identical to the ground truth in 17/33, while the rest (16 out 33) have an average deviation of 15.4\%. In Table~\ref{tab:comparison}, we consider \name{} results to be \emph{close to the ground truth} if the average deviation is less than 10\%.
Overall, these results show that \name{} is effective at executing the test suites of a large number of projects, and that the results are close to or equal to the ground truth in most cases.

To better understand the results, we analyze the projects by their main language (Figure~\ref{fig:effectivenessbyproject}).
For each language, we show the number of projects that fail to build, are successfully built, and have their test suites executed.
The results show that \name{} is most effective for Java, where it successfully builds and tests all projects.
The two languages that are the most difficult to handle are C and C++, which we attribute to the less standardized build and test processes in these languages, which often requires recompiling packages to be compatible with current system dependencies and project requirements.

Another perspective that we analyze is the presence of configurations for automated build tools in the project repositories. Among the 50 projects, 38 use automated build tools, such as Make and Cmake (in 14 projects), Npm and Yarn (13 projects), and Maven/Gradle (9 projects). \name{} successfully runs the tests of 28 out of the 38 projects that use automated build tools, while the approach succeeds for 5 out of the 12 projects not using such tools.
On the one hand, this result shows that the configuration and automation available within the project itself is a factor in the success of \name{}.
On the other side, the result also shows that \name{} is capable of handling projects that lack such automation.

We also study the presence and impact of submodules.
Submodules usually are standalone repositories that are included within the target project as a dependency and should be installed along with the target project.
Our dataset has 9 projects that use submodules. We find that \name{} struggles with such projects, as only 2 out of 9 reach a successful test suite execution.

\begin{table}
	\centering
	\caption{Effectiveness of \name{} and comparison to baselines.}
	\renewcommand{\arraystretch}{1} %
	\setlength{\tabcolsep}{7pt} %
	\small
	\begin{tabular}{@{}lrrrr@{}}
		\cline{2-5}
		& \textbf{ExecutionAgent} & \textbf{LLM scripts} & \textbf{AutoGPT} & \textbf{Flapy} \\ \hline
		Built      		& 41 / 50 & 29 / 50 & 9 / 50  & 6 / 10    \\ 
		Executed tests              & 33 / 50 & 5 / 50 & 4 / 50  & 0 / 10    \\ 
		Ground truth deviation < 10\%        & 29 / 50 & 4 / 50 & 2 / 50  & 0 / 10     \\ 
		\hline
	\end{tabular}
	\label{tab:comparison}
\end{table}

Finally, we compare the individual test case outcomes between \name{} and the ground truth. In 17 out of 33 tested projects, the same test sets were executed with matching outcomes. Among the remaining 16 projects, 99.9\% of passing test cases aligned with the ground truth, but deviation arose primarily in failing test cases. A detailed analysis of five projects with deviations revealed three key factors contributing to these differences. First, some projects are structured into independent modules with separate builds and test suites. For example, in the \textit{Activiti} project, a failure in compiling the \textit{bpmn} converter module halted the remaining testing process, though an existing option could have allowed the continuation of tests in other modules, which \name{} did not invoke. Second, incomplete dependency installations may cause test failures or errors. In the \textit{Axios} project, for instance, the absence of a headless Chromium binary resulted in an error, preventing 8 test cases from executing. Third, we attribute certain differences to test flakiness as errors occurred without clear indications of missing dependencies or setup issues.

\paragraph{Comparison with LLM scripts}
In Table~\ref{tab:comparison}, we compare \name{} to the three baselines.
The general-purpose, LLM-generated scripts successfully build many projects (29/50), but then often often fail to execute the test suites (only 5/29 succeed).
Inspecting the results, we find that the LLM scripts often do not account for the differences between the setup required by a regular user and the development setup.
The development setup often involves additional steps, such as installing additional software (e.g., a compiler, glibc-32, or a testing framework), recompiling some dependencies, or changing configuration files.
In contrast, the usage setup is much simpler, often using an already complete requirement/setup file, such as setup.py for Python and pom.xml for Java.
Because the LLM scripts do not account for these differences, they often fail to execute the test suite, whereas \name{} is able to iteratively fix unexpected errors.

\paragraph{Comparison with AutoGPT}
Even though AutoGPT is given the same budget as \name{}, it builds only 9/50 projects and successfully runs the tests suites for only 4 of them.
The results show that the task of running the tests of arbitrary projects is non-trivial, and that a specialized agent, such as \name{} is more effective at such tasks.
Fundamentally, \name{} differs from AutoGPT by using meta-prompting, by managing the memory of already performed commands more effectively, and by using more sophisticated and robust tools.

\paragraph{Comparison with Flapy}
The right-most column of Table~\ref{tab:comparison} compares with the Flapy baseline.
Because this baseline only targets Python, we apply it only to the 10 projects in our dataset that have Python as their main language.
The results show that Flapy reaches results similar to LLM-generated, general-purpose scripts in terms of building projects.
However, like the LLM scripts, it also struggles to execute the test suites, and cannot successfully complete the tests of any of the 10 projects.
Given the same 10 Python projects, \name{} successfully builds and tests 6 of them, all with testing results identical to the ground truth.

\subsection{Costs}
We evaluate the costs incurred by \name{} in terms of execution time and token usage.
The average time required by \name{} to process a project is 74 minutes (on a 256GB RAM Xeon(R) Silver 4214 machine with five projects being processed in parallel).
Given that the most reliable alternative to our approach is to manually set up and test the projects, we consider the time costs of \name{} to be acceptable.

\begin{figure}[t]
	\centering
	\begin{minipage}{0.5\textwidth}
	\centering
	\includegraphics[width=1\linewidth]{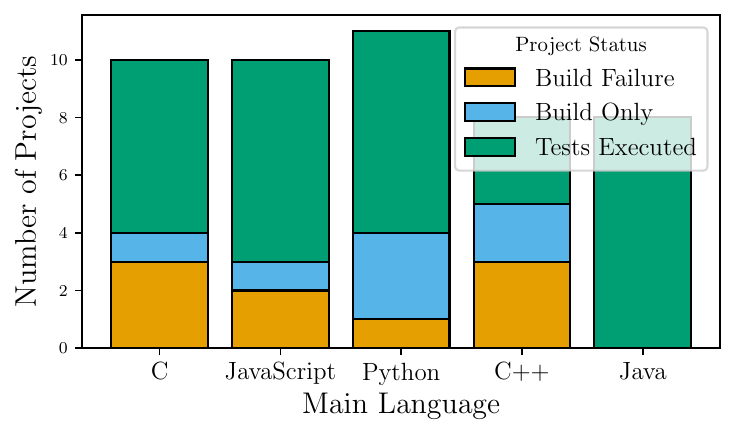}
	\caption{Effectiveness of \name{} by main language.}
	\label{fig:effectivenessbyproject}
	\end{minipage}\hfill
	\begin{minipage}{0.5\textwidth}
	\centering
	\includegraphics[width=0.96\linewidth]{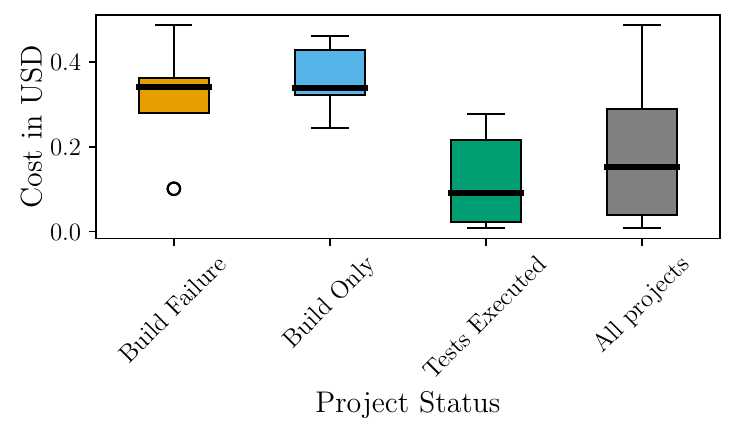}
	\caption[boxplot of costs]{Monetary costs due to LLM usage.}
	\label{fig:costsboxplot}
	\end{minipage}
\end{figure}

\name{} relies on an LLM, which incurs costs computed in terms of token usage.
Figure~\ref{fig:costsboxplot} shows the monetary costs due to the usage of the LLM.
Based on current pricing, the average cost of processing a project is USD~0.16.
The costs differs significantly depending on whether \name{} succeeds in building and testing the project.
For projects that fail to build or test, the costs are higher because \name{} tries to fix any issues until exhausting the budget, leading to an average cost of USD~0.34.
In contrast, for projects where the approach succeeds, the costs are lower, with an average of only USD~0.10.
Overall, we consider the current costs to be acceptable given the benefits of the approach, and expect costs to decrease over time as LLMs become more efficient and cheaper.

\subsection{Ablation Study}

To evaluate the contributions of different components of \name{}, we conduct an ablation study with multiple variants. The studied variants are as follows:

\begin{itemize}
	\item \textbf{No preparation phase:} The agent starts directly with the feedback phase, lacking the structured information from meta-prompting and retrieval.
	\item \textbf{No feedback phase:} Instead of iteratively refining its approach, the agent generates a single script using the information from the preparation phase.
	\item \textbf{No guidelines:} The agent does not use language-specific or general guidelines, reducing its ability to follow common installation and testing conventions.
	\item \textbf{No external retrieval:} No external information from web search or CI/CD scripts.
\end{itemize}

Table~\ref{tab:ablation} presents the results of these variants compared to the full approach. The ablated versions demonstrate significantly lower performance. Removing the preparation phase leads to a notable drop in success rates (12 successful builds and only 8 test executions), highlighting the importance of structured planning and context retrieval. The absence of feedback further reduces effectiveness, as it eliminates the iterative refinement process, leading to only 5 successful test executions. We learn that generated guidelines play a crucial role, as without them, the effectiveness drops by 50\%. While retrieving external information (e.g., CI/CD configurations and via web search) are important, removing it only drops the success rate by 21\% and increases costs by 13\%. 
Overall, these findings emphasize the importance of both the preparation phase and the feedback phase in ensuring successful project setup and testing.

\begin{table}
	\centering
	\caption{Ablation study with four variants of \name{}.}
	\renewcommand{\arraystretch}{1} %
	\setlength{\tabcolsep}{2pt} %
	\small
	\begin{tabular}{@{}lcccc@{}}
		\cline{2-5}
		\textbf{} & \textbf{Built} & \textbf{ | Executed tests} & \textbf{ | Deviation < 10\% | } & \textbf{Project cost (USD)} \\ \cline{1-5}
		ExecutionAgent        & 41 & 33 & 29 & 0.16 \\
		No preparation phase  & 12 & 8 & 7 & 0.19 \\
		No feedback phase     & 19  & 5 & 5 & 0.02 \\ 
		No guidelines     & 26 & 15 & 15 & 0.15 \\ 
		No CI/Web retrieval     & 31 & 24 & 23 & 0.18 \\ \cline{1-5} \\
	\end{tabular}
	\label{tab:ablation}
\end{table}

\subsection{Analysis and Discussion}

\subsubsection{Usage of Tools}
To understand the behavior of \name{}, we start by a quantitative analysis of its tool usage.
Overall, 74.8\% of calls made by the approach invoke the tool \code{terminal}, while 16.0\% and 8.2\% invoke \code{write\_file} and \code{read\_file}, respectively.
The ``end of task'' tool accounts for 1\% of all tool invocations.
This high-level analysis shows that \name{} uses all available tools, and that the terminal seems to be indispensible.

\begin{figure}
	\centering
	\includegraphics[width=0.8\linewidth]{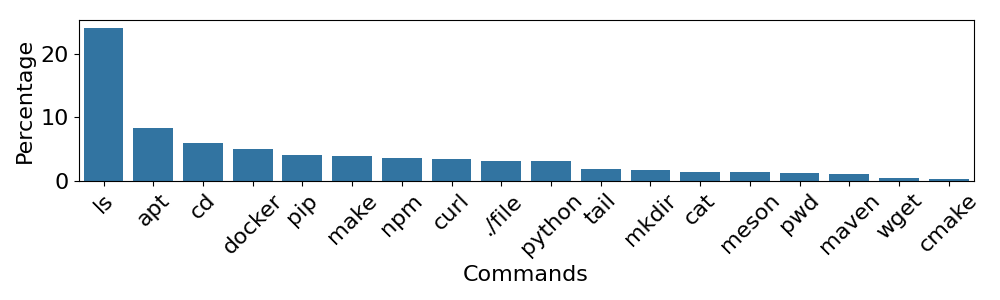}
	\caption{Distribution of most frequently used terminal commands (./file represents calls to local scripts).}
	\label{fig:terminalcmds}
\end{figure}

To further uncover the commands executed through the terminal, we count the frequency of all invoked commands.
Figure~\ref{fig:terminalcmds} shows the results for the top-10 commands.
We find that commands for exploring directories and files, such as \code{ls}, \code{cd}, and \code{pwd}, are common, summing up to 33\%.
Another subset of popular commands (21\%) is installation and build commands, such as \code{apt install} for system packages, \code{pip} for python and \code{npm} for JavaScript.
Finally, approximately 25\% of executed commands fall outside the top 10, demonstrating that \name{} utilizes a diverse command set, making a hard-coded selection of commands ineffective.

\subsubsection{Analysis of Trajectories}
We also perform a qualitative analysis of the trajectories of the agent, where ``trajectory'' refers to the sequence of steps that \name{} takes when dealing with a specific project.
The following describes three observations made during this analysis.

\paragraph{Recurring phases: Setup, dependencies, testing}
The trajectories of many projects, particularly software libraries, such as react-native, commons-csv, and pytest, include three clearly identifiable phases:
(1) \emph{Setup}: The agent typically starts with preparation commands, such as \code{ls}, followed by reading necessary files (e.g., \code{README.md}, \code{pom.xml}, or \code{package.json}) to understand the project structure.
(2) \emph{Dependency management}: Commands like \code{write\_to\_file} for creating Dockerfiles or setup scripts are prevalent. For instance, when creating a Dockerfile, the approach indicates the images to use, e.g., \code{FROM node:20} or \code{FROM python:3.10-slim}, followed by invoking additional commands to install dependencies using \code{apt-get} and package managers like \code{npm} or \code{yarn}.
(3) \emph{Execution of tests}: Subsequently, the approach issues commands to run tests (e.g., \code{npm test}, \code{pytest}) and then confirms that the tests were executed as expected.
Successful test execution is typically followed by logging of results, and by creating the scripts that \name{} is tasked to produce.

\paragraph{Consistent Docker usage}
Many trajectories involve using Docker to create isolated environments for applications (e.g., \code{docker build}, \code{docker run}). We observe two specific uses of Docker:
(1) \emph{Building and running containers}:
	For projects that provide their own Dockerfile, the agent often uses this Dockerfile to create an environment and install dependencies.
(2) \emph{Debugging and validating the environment:} 
	\name{} uses commands like \code{docker images} and \code{docker ps} to check that container images were created and to inspect running containers, showcasing an emphasis on ensuring the runtime environment is correctly configured.

\paragraph{Error handling and resilience}
\name{} demonstrates an ability to handle errors and unexpected outcomes effectively.
In particular, we observe the following strategies:
(1) \emph{Fallback actions:}
	Many commands in the produced scripts include fallback logic, such as using \code{||} to specify alternative actions when a command fails.
(2) \emph{Cleanup commands}:
	Commands are sometimes combined with cleanup actions to ensure that the environment is left in a consistent state, and that commands do not accumulate unnecessary files in case of errors.

\subsubsection{Complexity of Scripts Generated by \name{}}
We measure the script size (Section~\ref{s:metrics}) to assess the complexity of the scripts generated by \name{}. Across the 33 successful projects, the average script size is 18, with a minimum of 12 and a maximum of 37. That is, the final setup requires running 16 commands, on average. While this number may not seem particularly high, the scripts are concise because the agent creates them once it has found a successful sequence of commands.

\subsubsection{Limitations}
Through manual trajectory analysis, we identify two behaviors that frequently contribute to not succeeding in executing the tests.
First, the agent often repeats the same mistake, leading to wasted commands.
A common pattern is attempting to use \code{sudo} when it is unavailable, resulting in the agent correcting itself after an error, yet repeating the mistake in subsequent commands.
Second, \name{} often fails to follow up on certain commands. For instance, when installing a new version of Node or gcc, the old version frequently remains the default, unless explicitly changed. This oversight leads to repeated cycles of checking versions, installing new ones, and failing to set them as default, causing errors and wasting resources.

\section{Threats to Validity}
Our results are subject to several threats to validity.
First, \name{} typically runs tests in a single configuration (e.g., one language version, browser, or operating system), which may not capture variations across different environments, such as multiple browsers or OS versions. This limitation can be addressed by allowing users to modify the prompt to specify the desired configurations.
Second, the approach is designed around modern technologies (e.g., as asked for in the meta-prompt), which may limit the effectiveness for projects that require older dependencies. However, this limitation is mitigated by the technique's iterative approach, allowing it to detect and adapt to legacy dependencies when necessary.
Third, the approach has been primarily tested on popular projects with relatively good documentation. Results could differ on less well-known or poorly documented projects. In addition, a lot of projects in our dataset do not use submodules, which seem to challenge \name{}.
Finally, our criteria for selecting projects, especially the availability of ground truth test execution results, may bias the dataset toward projects that use some form of CI/CD.
Addressing this potential bias would require another way of establishing a ground truth, such as manually performing the task addressed by \name{} and comparing the results to those produced by the approach.

\section{Related Work}
\label{sec: rw}

\paragraph{Large language models in software engineering}

The field of software engineering has seen a rapid increase in the use of LLMs in recent years.
Generating code for a given function-level comment has become a standard task to evaluate the capabilities of LLMs~\cite{Chen2021,Liu2023a}.
To make code generation practical also for real-world software development, researchers have proposed techniques to augment prompts with repository-level context~\cite{Ding2022,Zhang2023a,Shrivastava2023,arXiv2024_De-Hallucinator}.
Beyond generating application code, LLMs have been used to generate unit tests~\cite{Lemieux2023,DBLP:journals/tse/SchaferNET24,Ryan2024,Kang2023,Feng2024,Pizzorno2024}, to translate code from one language into another~\cite{Roziere2020}, and to fuzz test programs that accept code as their input, such as compilers~\cite{icse2024-Fuzz4All}.
These approaches demonstrate the potential of LLMs to automate software development tasks that require understanding and generating code.
In addition to creating new code from scratch, LLMs have been used for modifying existing code.
One line of work focuses on predicting code edits based on previously performed edits~\cite{Wei2023,Gupta2023}.
Other work tries to automate common code changes, such as refactorings~\cite{Dilhara2024}.
A team at Meta shows how to use LLMs for augmenting existing, human-written tests~\cite{Alshahwan2024}.
Finally, there is work on multi-step code editing~\cite{Bairi2023}, where an LLM first plans multiple edit steps and then performs them one after the other.

Our work differs from all the above by focusing not directly on code, but on the task of setting up and running test suites of software projects.
We envision LLM-based techniques for code generation and code editing to benefit from our work by using the executable test suites as a feedback signal to evaluate the correctness of the generated code.

\paragraph{LLM-based agents}

Recent work has started to explore the power of LLM-based agents in software engineering tasks.
The most prominent agents focus on automated program repair, e.g., in RepairAgent~\cite{Bouzenia2024} and on automatically addressing issues that describe bugs, missing features, and other improvements of a code base, e.g., in SWE-Agent~\cite{Yang2024a}, MarsCode Agent~\cite{Liu2024a}, Magis~\cite{Tao2024}, and
AutoCodeRover~\cite{Zhang2024a}.
Another line of work explores an agent trying to describe the root cause of a software failure~\cite{Roy2024}.
We refer to a recent survey for a more comprehensive overview of recent work~\cite{Jin2024}.
To the best of our knowledge, our work is the first to explore the use of LLM-based agents for setting up and running test suites of software projects.

The concept of autonomous LLM agents is, of course, not limited to software engineering.
In 2022, researchers proposed to let LLMs generate and execute code to answer a given question~\cite{Gao2022}.
Building on this idea, others propose to augment LLMs with other tools invoked via APIs~\cite{Schick2023,Patil2023}.
The ReAct work~\cite{DBLP:conf/iclr/YaoZYDSN023} shows that LLM agents can outperform LLMs alone in a variety of tasks, e.g., question answering, fact verification, and interactive decision making tasks, such as webpage navigation.
Copra is an agent-based approach for formal theorem proving~\cite{thakur2024incontextlearningagentformal}. 
Two recent surveys give a comprehensive overview of such ``augmented LLMs''~\cite{Mialon2023} and LLM agents~\cite{Wang2023}.
Our evaluation compares to a general purpose LLM-based agent, AutoGPT, showing that \name{} outperforms it in the task of setting up and running test suites of software projects.

\paragraph{Benchmarks that rely on test suite executions}
Beyond helping developers, the ability to execute test suites is also essential for researchers.
In particular, several popular benchmarks rely on test suite executions, such as 
Defects4J~\cite{Just2014},
BugsInPy~\cite{widyasari2020bugsinpy},
SWE-bench~\cite{Jimenez2023}, and
DyPyBench~\cite{fse2024-DyPyBench}.
Such benchmarks are used for various purposes, e.g., to evaluate fault localization, automated program repair, and dynamic analyses.
\name{} could help to automate the creation of such benchmarks, by reducing the manual effort required to set up and run test suites.

\paragraph{Automated pipelines for test suite execution}
Several research projects explore automated pipelines to execute test suites at a larger scale, either to create benchmarks~, e.g., in BugSwarm~\cite{tomassi2019bugswarm} and GitBug-Java~\cite{silva2024gitbug}, or as part of an empirical study~\cite{Gruber023}.
These approaches each target one programming language, e.g., Java~\cite{tomassi2019bugswarm,silva2024gitbug} and Python~\cite{Gruber023}, and sometimes rely on projects using a specific CI/CD platform, e.g., TravisCI~\cite{tomassi2019bugswarm}.
Our evaluation empirically compares against language-specific baselines, showing that \name{} is more effective while supporting multiple languages.

\section{Conclusion}
In this paper, we introduced \name{}, an automated technique designed to address the complexities of test suite execution across diverse software projects. By leveraging a large language model-based agent, \name{} autonomously builds an arbitrary project and runs its tests, producing tailored scripts to streamline future test executions. Our approach is modeled after the decision-making processes of a human developer, using meta-prompting and iterative refinement to adapt to project-specific dependencies and configurations. An evaluation on 50 open-source projects using 14 languages demonstrates that \name{} successfully executes most test suites and closely matches ground truth results, with clear improvements over existing methods. With reasonable computational and financial costs, \name{} holds strong potential as a useful technique for developers, autonomous coding systems, and researchers needing reliable test execution capabilities across varied software ecosystems.

\section{Data Availability}
ExecutionAgent is publicly available at \url{https://github.com/sola-st/ExecutionAgent/} and \url{https://doi.org/10.5281/zenodo.15202434}.

\section{Acknowledgments}

This work was supported by the European Research Council (ERC, grant agreements 851895 and 101155832) and by the German Research Foundation within the DeMoCo and QPTest projects.

\bibliographystyle{ACM-Reference-Format}
\bibliography{references,referencesMP}

\end{document}